\documentclass[preprint,12pt]{elsarticle}
\usepackage{graphicx}
\usepackage{amssymb}
\usepackage{amsmath}

\begin{document}

\begin{frontmatter} 

\title{Neutrino Signals in Electron-Capture \\ Storage-Ring Experiments} 

\author{Avraham Gal} 
\address{Racah Institute of Physics, The Hebrew University, Jerusalem 91904, 
Israel} 

\date{\today} 

\begin{abstract} 
Neutrino signals in electron-capture decays of hydrogen-like parent ions $P$ 
in storage-ring experiments at the GSI Laboratory are reconsidered, with 
special emphasis placed on the storage-ring quasi-circular motion of the 
daughter ions $D$ in two-body decays $P\to D+\nu_e$. It is argued that, to the 
extent that daughter ions are detected, these detection rates might exhibit 
modulations with periods of order seconds, similar to those reported in the 
GSI storage-ring experiments for two-body decay rates \cite{GSI08,GSI13}. 
New dedicated experiments in storage rings, or using traps, could explore 
these modulations. 
\end{abstract}


\end{frontmatter}

\section{Introduction} 
\label{sec:intro} 

\begin{figure}[htb]  
\begin{center} 
\includegraphics[width=0.6\linewidth]{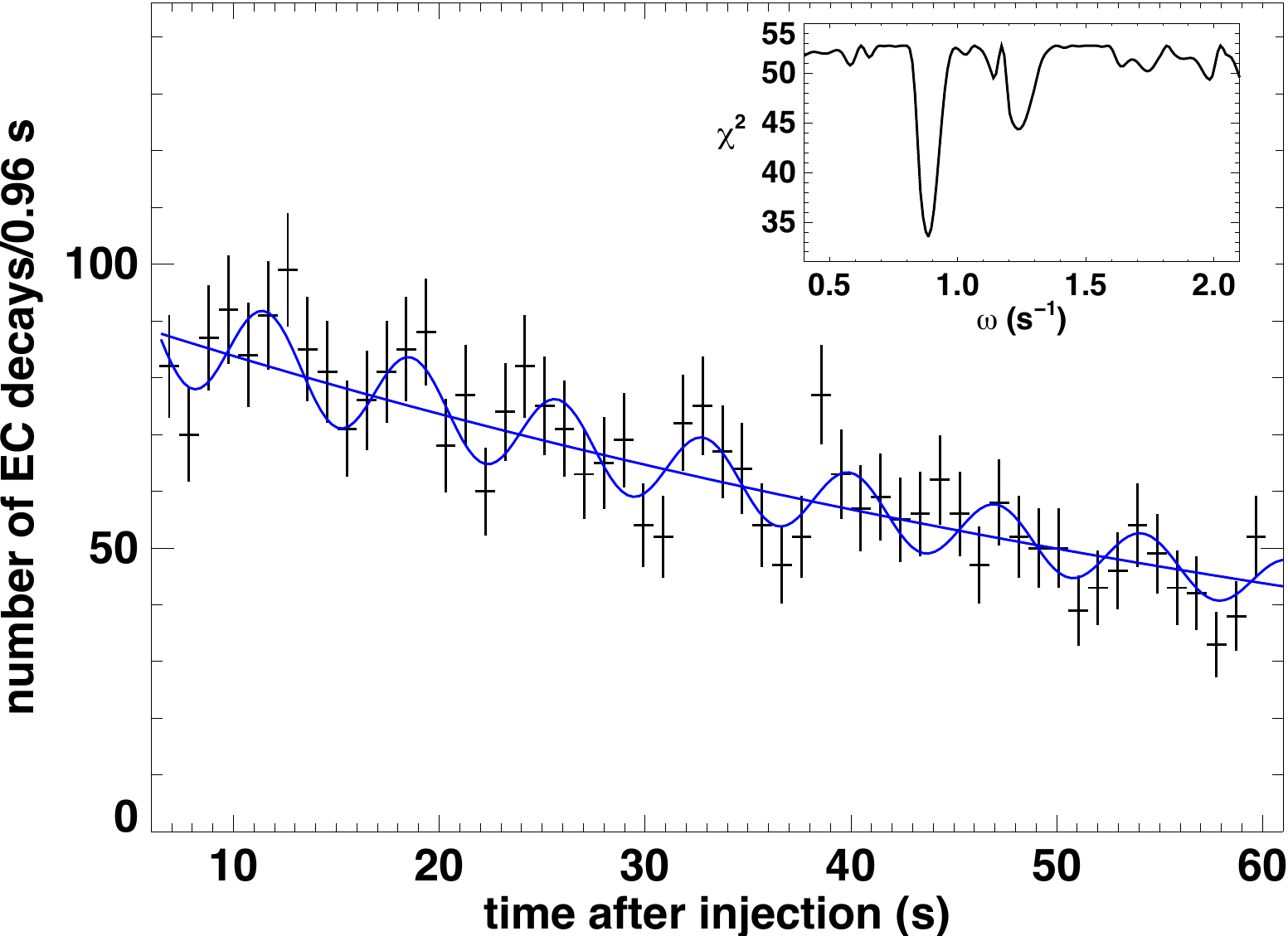} 
\caption{EC decay rate of H-like $^{142}$Pm$^{60+}$ ions circulating 
in the ESR of the GSI Laboratory, as recorded by a 245 MHz resonator. 
A decay-rate fit with angular frequency $\omega=0.884(14)$~s$^{-1}$, 
Eq.~(\ref{eq:omegaEC}), is also shown with $\chi^2$ values {\it vs.} 
$\omega$ given in the inset. Courtesy of Fritz Bosch 
(figure adapted from \cite{GSI13}).} 
\label{fig:142Pm} 
\end{center} 
\end{figure}

Electron capture (EC) rates for two-body decays $P\to D+\nu_e$ of 
hydrogen-like parent ions $P$ coasting in the GSI experimental storage 
ring (ESR) exhibit time modulation with laboratory period 
$\tau_{\rm GSI}\approx 7$~s~\cite{GSI08,GSI13} as shown for 
$^{142}$Pm in Fig.~\ref{fig:142Pm}. 
\noindent 
Here, the number of EC decays per time interval $dt$ is given, in obvious 
notation, by 
\begin{equation} 
\frac{dN_{\rm EC}(t)}{dt}=N(0)\, \exp(-\lambda t)\, {\lambda_{\rm EC}}\, 
\left ( 1+A\,\cos (\omega_{\rm EC}\,t+\phi)\right ),
\label{eq:omegaEC} 
\end{equation} 
where $t$ = 0 marks the time at which the parent ions $P$ were injected to 
the ESR. The modulation angular frequency $\omega_{\rm EC}$ corresponds 
to a minute energy splitting $\hbar\omega_{\rm EC} \approx 0.83\times 
10^{-15}$~eV in the parent-ion rest frame. The value of its amplitude is 
given by $A$ = 0.107 $\pm$ 0.024~\cite{GSI13}, half of that in the preceding 
experiment \cite{GSI08}. The decay of parent ions in the GSI experiments is 
signaled, to a variable degree, by the correlated observation of daughter 
ions which are also confined to a storage-ring motion, as discussed below. 
In contrast to these storage-ring results, non storage-ring experiments 
elsewhere found no trace of EC decay-rate modulations \cite{LBL08,TUM09}. 
A new storage-ring experiment, aimed at confirming the reported modulation, 
has been run recently at \mbox{GSI \cite{GSI14}}. 

Several works, beginning with the first GSI report \cite{GSI08}, noticed that 
the mass-eigenstate neutrino energies $E_{\nu_j}(p_j)$ in the $P\to D+\nu_j$ 
EC decay differ in the parent-ion rest frame P by 
\begin{equation} 
|E^{(P)}_{\nu_2}(p_2)-E^{(P)}_{\nu_1}(p_1)|=\frac{\Delta{m_{\nu}^2}}{2M_P}=
0.29\times 10^{-15}~{\rm eV} \approx \frac{1}{3}\hbar\omega_{\rm EC}, 
\label{eq:naive} 
\end{equation} 
suggesting perhaps that the reported modulation arises from interference 
between EC amplitudes $A_{\nu_j}(P\to D+\nu_j;~t)$ that encode quantum 
entanglement of the daughter ion with the mass-eigenstate neutrinos $\nu_j$ 
($j=1,2$) to which the emitted (but undetected) neutrino $\nu_e$ is dominantly 
coupled. In Eq.~(\ref{eq:naive}), $\Delta{m_{\nu}^2}\equiv |m_2^2-m_1^2|=
(7.6\pm 0.2)\times 10^{-5}~{\rm eV}^2$ is the squared-mass difference 
of these two mass-eigenstate neutrinos \cite{PDG} and $M_P\approx 132$~GeV 
is the rest mass of $^{142}$Pm. We note that the natural scale for 
neutrino-oscillation frequencies $\omega_{\nu}$ is $\hbar\omega_{\nu}=
\Delta{m_{\nu}^2}/2E_{\nu}$ which, for $E_{\nu}\sim4$~MeV, is about four 
orders of magnitude larger than the reported $\hbar\omega_{\rm EC}$. Ivanov 
and Kienle, in particular, attempted to derive rigorously $\hbar\omega_{
\rm EC}$ as the underlying energy splitting in two-body EC decays from 
such interference \cite{IK09}. However, since mass-eigenstate neutrinos 
$\nu_j$ are distinct particles, and neutrinos remain undetected in the GSI 
storage-ring experiments, the general rules of quantum mechanics require to 
sum up the amplitudes $A_{\nu_j}$ incoherently, $|A_{\nu_1}|^2+|A_{\nu_2}|^2$, 
with no amplitude interference terms arising in the calculated EC decay 
rate.{\footnote{However, if neutrinos are Dirac fermions with magnetic 
moment $\mu_{\nu}$, the amplitude for neutrinos $\nu_j$ of a given mass 
eigenstate that precess upon production in the magnetic field $\vec B$ of 
the storage ring interferes with that for the same mass-eigenstate neutrinos 
$\nu_j$ that are produced at regions of the storage ring unexposed to the 
magnetic field. Both amplitudes lead to the same left-handed $\nu_j$ final 
state \cite{Gal10}. To explain the reported GSI modulation by such 
interference, the value of the neutrino magnetic moment should be 
$\mu_{\nu_e}\approx 0.9\cdot 10^{-11} \mu_B$, where $\mu_B$ is the Bohr 
magneton. This value is about six times {\it lower} than the laboratory 
upper limit from Borexino \cite{Borexino08}.}} This point was forcefully 
made in Refs.~\cite{Giunti08,Kienert08,CGL09,Merle09,Flam10,Wu10,Pesh14}. 
\mbox{A similar argument} for the impossibility to have amplitude 
interference, and hence time modulations in daughter-ion {\it appearance} 
rates, is briefly reviewed in Sect.~\ref{subsec:standard}. 

\begin{figure}[htb]  
\begin{center}  
\includegraphics[width=0.7\linewidth]{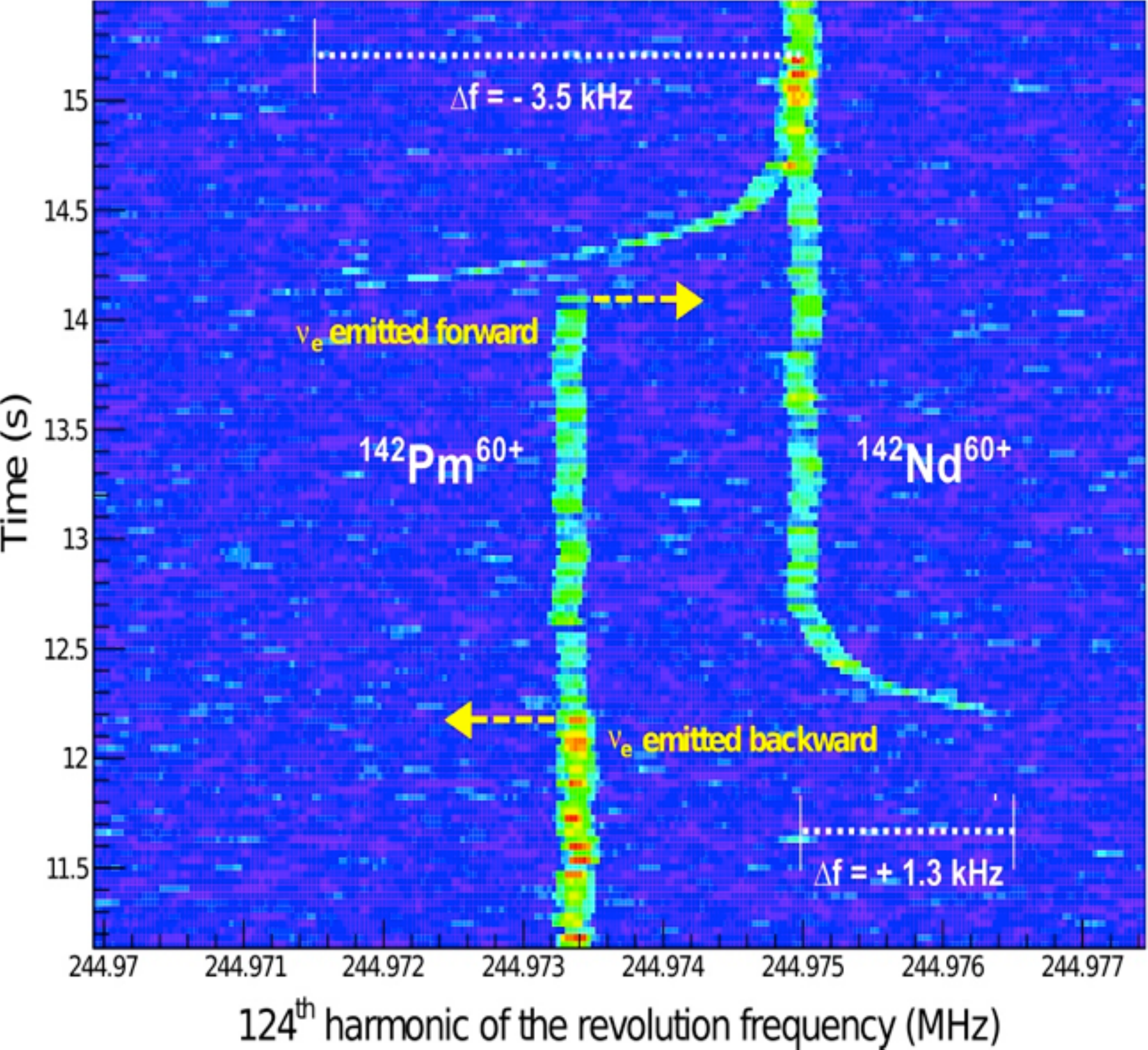} 
\caption{Traces of two cooled $^{142}$Pm$^{60+}$ parent ions {\it vs.} the 
time (in s) since injection, with time and frequency binning of 32 ms and 
31.25 Hz, respectively. Both parent H-like $^{142}$Pm$^{60+}$ ions decay by EC 
to bare $^{142}$Nd$^{60+}$ daughter ions, with arrows indicating the true decay 
times, as identified by a decrease of the intensity of the trace of the parent 
ions and the simultaneous onset of the trace of the recoiling daugher ion. The 
latter starts at a revolution frequency shifted by $\Delta f$ with respect to 
the frequency after completion of electron cooling. Courtesy of Fritz Bosch 
(figure adapted from \cite{GSI13}).} 
\label{fig:bosch1_14} 
\end{center} 
\end{figure}

The present work is focused on studying the time evolution of daughter ions 
$D$ in the GSI storage-ring spatially confined quasi-circular motion. Here, 
the storage-ring magnetic field disentangles momentarily the daughter ion 
from the neutrino with which it was produced in the EC decay. At subsequent 
times the daughter ion and the neutrino are no longer entangled together by 
the requirement that their total linear momentum vanishes in the parent rest 
frame P. The quasi-circular motion of $D$ is not correlated with the outgoing 
neutrino rectilinear path, and it is legitimate to treat it independently of 
the emitted neutrino so long as the remaining constraints imposed by the EC
decay kinematics are respected. Furthermore, we note that the GSI technique 
of recording parent decay events requires the unambiguous identification of 
the daughter ions over times of order 0.5~s, during which they coast in the 
storage ring for a total of $\sim$10$^6$ revolutions while their momentum 
spread is reduced drastically by the applied electron cooling, as shown 
in Fig.~\ref{fig:bosch1_14} for two daughter-ion frequency trajectories. 
A key question is whether or not the coherence prevailing at the time of the 
daughter-ion production could be sustained during the short period before the 
applied electron cooling might have completely wiped it out. Some comments on 
this issue, focusing on the role of electron cooling in the GSI storage-ring 
EC decay experiments \cite{NIMA04}, are made in Sect.~\ref{subsec:lab}.

Provided the answer to the question posed above is positive, meaning that 
some coherence is sustained, it is shown in Sects.~\ref{subsec:SR} and 
\ref{subsec:phases} that the time evolution of daughter ions in their 
storage-ring quasi-circular confined motion gives rise to modulations with 
angular frequency close to that given in (\ref{eq:naive}) and thereby close 
to that reported by the GSI experiments. Such modulations do not arise in 
rectilinear two-body decays.{\footnote{For a prototype of two-body decays, 
$\pi^+\to\mu^++\nu_{\mu}$, beginning with Ref.~\cite{Dolgov97} it has been 
concluded that the time evolution of recoil muons moving rectilinearly does 
not exhibit oscillations; see also Ref.~\cite{BK12}.}} This aspect of the GSI 
storage-ring experiments is considered here for the first time. Our prediction 
may be tested by keeping track of the steady motion of daughter ions in the 
storage ring under stable electron cooling for as long as possible. In this 
way, the weight with which daughter ions affect the decay-rate measurement 
could be varied and thereby evaluated.

\section{Daughter-Ion Oscillations in Two-Body EC} 
\label{sec:2bodyEC} 

In the following subsections we confront the standard case in which the 
daughter-ion motion is not subject to any external electromagnetic (EM) 
field with that of a daughter-ion quasi-circular storage-ring motion guided 
by external EM fields.

\subsection{Daughter-Ion Oscillations in Two-Body EC: Non Storage-Ring Motion} 
\label{subsec:standard} 

To discuss the disentanglement of a two-body system created at time $t=0$ 
in EC decay and moving under no external constraints, we briefly recover 
here the argumentation given in Ref.~\cite{Kienert08}. The product state 
$|\Psi_{D+\nu_e}(t)\rangle$ of $D$ plus electron neutrino $\nu_e$ entangled 
pair is expanded in the mass-eigenstate neutrino particle basis 
$\nu_j,~j=1,2,3$, 
\begin{equation} 
|\Psi_{D+\nu_e}(t)\rangle = \sum_j U_{ej}\,|\psi_{D+\nu_j}(t) \rangle,
\label{eq:Psi} 
\end{equation} 
where $U_{ej}$ are elements of the unitary matrix $U$ connecting the electron 
(flavor) neutrino $\nu_e$ with mass-eigenstate neutrinos $\nu_j$. The 
state vector $|\psi_{D+\nu_j}(t)\rangle$ describes the space-time motion 
of the daughter ion $D$ and mass-eigenstate neutrino $\nu_j$ which are 
entangled through momentum conservation: ${\vec p}_{Dj}+{\vec p}_{\nu_j}=0$ 
in the parent-ion rest frame P. For simplicity we suppressed the spatial 
dependence everywhere. The density matrix $\rho_{\Psi}(t)$ associated with 
$|\Psi_{D+\nu_e}(t)\rangle$ is given by 
\begin{equation}
\rho_{\Psi}(t) = |\Psi_{D+\nu_e}(t)\rangle \langle \Psi_{D+\nu_e}(t)|.
\label{eq:rho_Psi}
\end{equation} 
The observation of the daughter ion $D$ at a given time $t>0$ disentangles the 
two-body system. This requires to project out the undetected neutrino. To this 
end one defines an appropriate \mbox{projection operator }
\begin{equation} 
Q_{\nu}=\sum_j\int{d^3p_{\nu}|\psi_{D+\nu_j},{\vec p}_{\nu}\rangle\langle
\psi_{D+\nu_j},{\vec p}_{\nu}|},  
\label{eq:Q} 
\end{equation} 
where the sum and integral run over a complete set of mass-eigenstate 
neutrinos $\nu_j$ with momenta ${\vec p}_{\nu}$. The summation on $\nu_j$ 
could in principle be replaced by summing on flavor neutrinos $\nu_{\alpha}$, 
demonstrating that $Q_{\nu}$ is flavor blind. Taking the trace of $Q_{\nu}
\rho_{\Psi}(t)$ in the $\nu_j$ basis and also integrating over the neutrino 
phase space gives 
\begin{equation}
{\rm Tr}_{\nu}(Q_{\nu}\rho_{\Psi}(t)) = \sum_j|U_{ej}|^2 = 1, 
\label{eq:trace1}
\end{equation}
where the orthogonality of mass-eigenstate neutrinos, 
$\langle\nu_j|\nu_{j'}\rangle = \delta_{jj'}$, was instrumental to eliminate 
interference terms that could have led to daughter ion $D$ oscillations.

\subsection{Daughter-Ion Oscillations in Two-Body EC: Storage-Ring Motion I} 
\label{subsec:SR} 

As argued in the Introduction, the external magnetic field that confines 
$D$ in its circular motion disentangles $D$, for each of its equal-mass 
energy-momentum components $D_j$, from the propagating mass-eigenstate 
neutrinos $\nu_j$. This occurs momentarily upon EC decay, following 
which the space-time motion of the daughter ion $D$ and mass-eigenstate 
neutrinos $\nu_j$ is no longer entangled through momentum conservation: 
${\vec p}_{Dj}+{\vec p}_{\nu_j}=0$. Detection of $D$ occurs at a later time. 
It is justified then to focus on the storage-ring motion of $D$, disregarding 
the undetected neutrino trajectory except through the dependence of the 
circular-motion temporal phase $E_{Dj}\tau$ (see below) on the $\nu_j$ 
neutrino label $j$. Here, $\tau=t-t_d$ denotes time $t$ with respect to 
the EC decay time $t_d$. To disentangle, we first project the state vector 
$|\Psi_{D+\nu_e}(t)\rangle$, Eq.~(\ref{eq:Psi}), on the produced electron 
neutrino state vector $|\nu_e(t)\rangle$ in the limit $t\to t_d$, and 
propagate the $D$ initial state vector $|\Psi_{D}(t_d)\rangle$ forward in 
time, \mbox{thereby obtaining} 
\begin{equation} 
|\Psi_{D}({\vec x},~t)\rangle = {\cal N}_{D} \sum_{j}|U_{ej}|^2\,|\psi_{Dj}
({\vec x},~t)\rangle,  
\label{eq:BK22a}
\end{equation} 
where the spatial ($\vec x$) dependence was made explicit and ${\cal N}_{D}$ 
is a space-time independent normalization constant. The state-vector 
$|\psi_{Dj}\rangle$ wave packet $\psi_{Dj}$ is given, in a plane-wave limit, 
by 
\begin{equation} 
\psi_{Dj}({\vec x},~t)=\exp [i({\vec p}_{Dj}\cdot {\vec\xi}-E_{Dj}\tau)] 
\label{eq:BK6} 
\end{equation} 
per each mass-eigenstate neutrino $\nu_j$ with which $D$ was entangled at the 
EC decay time $t_d$. Here, ${\vec\xi}={\vec x}-{\vec x}_d$ and $\tau=t-t_d$. 
Using a more realistic wave packet rather than a plane wave for $\psi_{Dj}$ 
does not change the nature of conclusions reached below. To obtain the 
amplitude $A_D$ for detecting $D$ at time $t$, subsequent to its production 
at time $t_d$ in the two-body EC decay $P\to D+\nu_e$, we form the matrix 
element between the $D$ state vector $|\Psi_{D}({\vec x},~t)\rangle$ and 
the initial state vector $|\Psi_{D}({\vec x}_d,~t_d)\rangle$, normalizing 
it so that $A_D(t$=$t_d)$=1. The amplitude $A_D$ generated in this way is 
essentially identical with Eq.~(\ref{eq:BK22a}): 
\begin{equation} 
A_D=\sum_{j=1,2,3}|U_{ej}|^2\psi_{Dj}({\vec x},~t).  
\label{eq:BK22} 
\end{equation} 
The space-time wave packets $\psi_{Dj}({\vec x},~t)$ in (\ref{eq:BK22}) 
are weighed by the {\it probabilities} $|U_{ej}|^2$ with which each $D_j$ 
was produced at time $t_d$ of the EC decay. The use of probabilities, 
$\sum_{j=1,2,3}|U_{ej}|^2=1$, rather than amplitudes $U_{ej}$ or their 
complex conjugates $U^{\ast}_{ej}$, is consistent with the assumption that the 
daughter ion $D$ was disentangled momentarily by the action of the external 
magnetic field from the electron neutrino $\nu_e$ with which it had been 
entangled at the moment of EC decay. Forming the probability $|A_D|^2$ 
gives rise to interference terms $\psi^{\ast}_{Di}\psi_{Dj}$, $i\neq j$, 
that generate oscillations, unless these interference terms come out 
independent of space and time as it happens in rectilinear motion; 
see below.  

The appearance of the probabilities $|U_{ej}|^2$ in Eq.~(\ref{eq:BK22}) is 
reminiscent of a $\nu_e\to\nu_e$ amplitude for neutrino detection with no 
flavor change \cite{BK12}. To check that the corresponding amplitude, 
\begin{equation} 
A_{\nu_e}=\sum_{j=1,2,3}|U_{ej}|^2\psi_{\nu_j}({\vec x},~t),
\label{eq:BKnu} 
\end{equation} 
is also obtainable in the present formalism one projects the state vector 
$|\Psi_{D+\nu_e}(t)\rangle$, Eq.~(\ref{eq:Psi}), on the daughter-ion 
initial state vector $|\Psi_{D}(t_d)\rangle$, resulting in 
\begin{equation} 
|\Psi_{\nu}({\vec x},~t)\rangle = {\cal N}_{\nu} \sum_{j}U_{ej}\,|\psi_{\nu_j}
({\vec x},~t)\rangle.  
\label{eq:BK22b}
\end{equation} 
Forming the matrix element between the $\nu$ state vector $|\Psi_{\nu}({
\vec x},~t)\rangle$ and the state vector $|\Psi_{\nu_e}({\vec x}_d,~t_d)
\rangle$, and normalizing so that $A_{\nu_e}(t$=$t_d)$=1, one obtains the 
amplitude $A_{\nu_e}$ in the form Eq.~(\ref{eq:BKnu}), as anticipated for it.

\subsection{Daughter-ion Oscillations in Two-Body EC: Storage-Ring Motion II} 
\label{subsec:phases} 

Adopting the idealized plane wave form (\ref{eq:BK6}) in the parent-ion 
rest frame P, and assuming rectilinear motion of the stable daughter 
ion $D$ produced in the two-body decay $P\to D+\nu_e$, it was shown in 
Ref.~\cite{BK12} that to order ${\cal O}(\Delta{m_{\nu}^2})$ the $j$ 
dependence of the spatial phase ${\vec p}_{Dj}\cdot{\vec\xi}$ cancels 
the $j$ dependence of the temporal phase $E_{Dj}\tau$. This leads to 
$\psi^{\ast}_{Di}\psi_{Dj}=1$ for all $i$ and $j$. Owing to Lorentz covariance 
of the space-time phase in (\ref{eq:BK6}), $\psi_{Dj}({\vec x},~t)$ is to 
an excellent approximation independent of $j$ for rectilinear motion in any 
inertial coordinate frame boosted from the parent-ion rest frame P. Hence, 
a rectilinearly outgoing daughter-ion $D$ should incur no modulation or 
oscillation. 

The GSI experiment differs in one essential respect from EC experiments 
done with rectilinear geometry \cite{LBL08,TUM09}. In the GSI experiment 
the daughter ion traverses a storage-ring quasi-circular trajectory. In this 
case, because $D$ is confined to the storage ring, it can be shown that 
interference effects arising from the phases ${\vec p}_{Dj}\cdot {\vec\xi}$ 
are negligible although each of these phases need not be particularly small. 
Thus, recalling that 
\begin{equation} 
p^{(P)}_{Dj}-p^{(P)}_{Dj'}=p^{(P)}_{{\nu}_{j'}}-p^{(P)}_{{\nu}_j}=
\frac{\Delta{m_{jj'}^2}}{2E_{\nu}}\;\;\;\;\; 
(\Delta{m_{jj'}^2}\equiv m_j^2-m_{j'}^2) 
\label{eq:p} 
\end{equation} 
in the parent-ion rest frame P, with the same order of magnitude in the 
laboratory frame L, and using a spatial size of $\xi\sim 50$~m, the {\it 
difference} of two such phases is negligible, $|(p_{Dj}-p_{Dj'})\xi|{\ll} 1$, 
and will be disregarded henceforth in considering potential contributions 
to interference terms. This result follows also upon replacing the plane 
wave (\ref{eq:BK6}) by a wave packet that reflects the storage ring confined 
motion. Suppressing therefore ${\vec p}_{Dj}\cdot {\vec\xi}$ spatial phases, 
a nonvanishing $j$ dependence emerges from the phase $E_{Dj}\tau$. Assuming 
that no further $j$ dependence arises from the gone-away neutrino, and using 
(\ref{eq:naive}), interference terms between $\psi_{Dj}(x\approx 0,t)$ and 
$\psi_{Dj'}(x\approx 0,t)$ for $j\neq j'$ are given in the parent-ion rest 
frame P by 
\begin{equation} 
2\cos(\Delta E^{(P)}_{Djj'}\tau_P)=2\cos(\frac{\Delta{m_{jj'}^2}}{2M_P}\tau_P) 
\label{eq:interf} 
\end{equation} 
where the rest-frame time $\tau_P=0$ corresponds to the EC decay time, and 
\begin{equation} 
\Delta E^{(P)}_{Djj'}\equiv E^{(P)}_{Dj}-E^{(P)}_{Dj'}=
E^{(P)}_{\nu_{j'}}(p_{{\nu}_{j'}})-E^{(P)}_{\nu_j}(p_{{\nu}_j})=
\frac{\Delta{m_{j'j}^2}}{2M_P}, 
\label{eq:E}
\end{equation} 
see also Eq.~(\ref{eq:naive}). We note that these interference terms 
are evaluated at a given space-time point $(x\approx 0,t)$, regardless of 
any space-time path that might have been used to lead to the amplitude 
$A_D$ of Eq.~(\ref{eq:BK22}). The angular frequency suggested by 
Eq.~(\ref{eq:interf}) for daughter-ion oscillations comes close 
indeed to the GSI value $\omega_{\rm EC}$. A more rigorous derivation 
of this rest-frame result is provided in Sect.~\ref{subsec:cm}, 
and the transition to the lab frame is detailed in Sect.~\ref{subsec:lab}. 
Unlike transformations of decay rates which require only the well tested 
$t_L=\gamma t_P$ relationship between the lab time $t_L$ and the proper time 
$t_P$ \cite{Giunti14}, a more careful consideration of Lorentz transformations 
involving momentum-energy and space-time is needed here.

\section{Results} 
\label{sec:results} 

\subsection{center-of-mass Oscillation Expressions} 
\label{subsec:cm} 

As argued in the Introduction, the external magnetic field that confines 
$D$ to a circular motion and the application of electron cooling in fact 
disentangle $D$, for each of its energy-momentum components $D_j$, from 
the propagating mass-eigenstate neutrinos $\nu_j$. This allows us to focus 
on the storage-ring motion of $D$, disregarding the undetected neutrino 
trajectory except implicitly through the dependence of the circular-motion 
phase $E_{Dj}\tau$ in (\ref{eq:BK6}) on the $\nu_j$ neutrino label $j$. 
Consider then the amplitude $A_D$ (\ref{eq:BK22}) for detecting $D$ at time 
$t$ subsequent to its production at time $t_d$ in the two-body EC decay $P\to 
D+\nu_e$. Recalling that $\psi_{Dj}({\vec x},~t)$, Eq.~(\ref{eq:BK6}), reduces 
to $\exp(-iE_{Dj}^{(P)}\tau)$, $A_D$ is given in the parent-ion rest frame P 
by 
\begin{equation} 
A_D=\sum_{j=1,2,3}|U_{ej}|^2 \exp(-iE_{Dj}^{(P)}\tau_P), \;\;\; 
\tau=t-t_d \geq 0, 
\label{eq:BK22final} 
\end{equation} 
where $U_{ej}$ are elements of the unitary matrix $U$ connecting the electron 
(flavor) neutrino $\nu_e$ with mass-eigenstate neutrinos $\nu_j$. Specializing 
to a $2\times 2$ orthogonal neutrino mixing matrix with angle $\theta$, the 
relevant probability $|A_D|^2$ is given by 
\begin{equation} 
|A_D|^2=1-{\sin}^2(2\theta)\,{\sin}^2(\frac{\Delta E^{(P)}_{D12}}{2}\,
\tau_P),  
\label{eq:BK35}
\end{equation} 
which exhibits the same oscillation angular frequency $\Delta E^{(P)}_{D12}$ 
discussed in the previous subsection. The structure of Eq.~(\ref{eq:BK35}) 
coincides, fortuitously though, with the well known neutrino deficit formula 
for $|A_{\nu_e}|^2$, where $A_{\nu_e}$ is defined by Eq.~(\ref{eq:BKnu}), in 
a $\nu_e$ disappearance experiment. Equation (\ref{eq:BK35}), which assumes 
a given parent-ion decay time $t_d$, or equivalently daughter-ion appearance 
time $t_d$, does not yet give the differential rate for EC decay marked 
by detection of the daughter ion. One needs first to let $t_d$ vary 
between injection time $t_d=0$ ($\tau_P=t_P$) to detection time $t_d=t$ 
($\tau_P=0$).{\footnote{At this point we disregard the exponential intensity
depletion of the decay ion at time $t_d$ which leads, in a more systematic
derivation, to the appearance of the familiar $\lambda\,\exp(-\lambda t_P)$
prefactor in all of our subsequent $D$ detection rates.}} Hence, we integrate 
over $\tau_P$ between 0 to $t_P$ and then differentiate with respect to $t_P$ 
to get the required $D$ detection rate. This results in precisely the same 
expression for the daughter-ion detection rate as Eq.~(\ref{eq:BK35}): 
\begin{equation} 
R_D(t_P)=|A_D(t_P)|^2=1-{\sin}^2(2\theta)\,{\sin}^2
(\frac{\Delta E^{(P)}_{D12}}{2}\,t_P),  
\label{eq:BKextra} 
\end{equation} 
with $\Delta E^{(P)}_{D12}=\Delta{m_{\nu}^2}/2M_P$, see Eq.~(\ref{eq:interf}) 
and the text that follows it. Recall that $\Delta E^{(P)}_{D12}$ was obtained 
for storage-ring confined motion of $D$. As discussed in the previous 
subsection, the phase $\Delta E^{(P)}_{D12}\tau_P$ gets canceled in 
rectilinear motion by a spatial contribution that is suppressed in the 
GSI storage-ring motion. Therefore, to extrapolate the storage-ring rate 
(\ref{eq:BKextra}) to EC rectilinear motion, one substitutes formally 
$\Delta E^{(P)}_{D12}=0$, obtaining thereby $R_D(t_P)_{\rm rectilinear}=1$, 
in agreement with standard discussions of recoil oscillations \cite{BK12}. 
Expression (\ref{eq:BKextra}) exhibits modulations with amplitude 
\begin{equation} 
A=\frac{{\sin}^2(2\theta)/2}{1-{\sin}^2(2\theta)/2}\approx 0.733\pm 0.021 
\label{eq:ampl} 
\end{equation} 
relative to a base line of $1-{\sin}^2(2\theta)/2\approx 0.577\pm 0.011$,
where the value ${\sin}^2(2\theta)=0.846\pm 0.021$ \cite{PDG} was used. 
This summarizes our prediction for future experiments focusing on 
daughter-ion oscillations in confined motion.

\subsection{Laboratory Frame Oscillation Expressions} 
\label{subsec:lab} 

In the laboratory frame L, the plane wave (\ref{eq:BK6}) assumes the form 
\begin{equation} 
\psi^{(L)}_{Dj}(\vec x_L,~t_L)=\exp [i({\vec p}^{\,(L)}_{Dj}\cdot{\vec\xi_L}
-E^{(L)}_{Dj}\tau_L)],  
\label{eq:lab}
\end{equation}  
which upon suppressing the spatial phase ${\vec p}^{\,(L)}_{Dj}\cdot
{\vec\xi_L}$ gives rise to a lab interference term 
\begin{equation} 
2\cos(\Delta E^{(L)}_{Djj'}\tau_L), 
\label{eq:interfL} 
\end{equation}
with oscillation angular frequency that corresponds to a lab energy splitting 
$\Delta E_{Djj'}^{(L)}\equiv E^{(L)}_{Dj}-E^{(L)}_{Dj'}$. This is evaluated 
by transforming from the rest frame P to the lab frame L, 
\begin{equation} 
\Delta E_{Djj'}^{(L)}=\gamma\Delta E_{Djj'}^{(P)}+\gamma{\vec\beta}\cdot
\Delta {\vec p}_{Djj'}^{\,(P)}, 
\label{eq:DelE_L1} 
\end{equation} 
where $\Delta {\vec p}_{Djj'}^{\,(P)}\equiv {\vec p}_{Dj}^{\,(P)}-{\vec p}_
{Dj'}^{\,(P)}$ and $\beta=0.71$ with $\gamma\equiv (1-\beta^2)^{-1/2}=1.42$ 
\cite{GSI08,GSI13}. This Lorentz transformation holds everywhere on the 
storage-ring trajectory of $D$. Recalling from (\ref{eq:p}) and (\ref{eq:E}) 
that $|\Delta E_{Djj'}^{(P)}/\Delta p_{Djj'}^{(P)}|=E_{\nu}/M_P\ll 1$, one 
could argue that the first term on the right-hand side of (\ref{eq:DelE_L1}) 
may be safely neglected with respect to the second one which by (\ref{eq:p}) 
is of order $\Delta{m_{\nu}^2}/2E_{\nu}$, several orders of magnitude larger 
than $\hbar\omega_{\rm EC}$. This would render unobservable the oscillations 
of daughter ions coasting in the GSI storage ring. At this point we recall 
that the electron cooling applied at the GSI storage ring \cite{NIMA04} 
primarily affects the longitudinal momentum of $D$, parallel to the 
beam direction $\hat{\beta}$, through small-angle Coulomb scattering, as 
demonstrated in Fig.~\ref{fig:bosch1_14} for the two trajectories depicted for 
$D$. This is likely to cause partial decoherence by removing the distinction 
between the separate longitudinal components $p_{\parallel Dj}$. We therefore 
impose the condition ${\vec\beta}\cdot\Delta{\vec p}_{Djj'}^{\,(P)}=0$, 
resulting thereby in 
\begin{equation} 
\Delta E_{Djj'}^{(L)}=\gamma\Delta E_{Djj'}^{(P)}=
\frac{\gamma\Delta{m_{j'j}^2}}{2M_P}. 
\label{eq:DelE_L2} 
\end{equation} 
Note that $\Delta p_{Djj'}^{(L)}$ too comes out as small, 
$\Delta p_{Djj'}^{(L)}=\beta\gamma\Delta{m_{jj'}^2}/2M_P$, upon arguing 
similarly to the argumentation applied above. 

Substituting $\Delta E_{Djj'}^{(L)}$ from (\ref{eq:DelE_L2}) in the 
interference term (\ref{eq:interfL}) leads to the following period 
of daughter-ion oscillations in the lab frame L: 
\begin{equation} 
T^{(L)}_{\rm osc}=\frac{4\pi\;M_P}{\gamma\Delta(m_{\nu}^2)}=10.17\pm 
0.26~{\rm s}, 
\label{eq:Tosc} 
\end{equation} 
where the uncertainty is due to the uncertainty in the value of 
$\Delta(m_{\nu}^2)$. Compared with the reported value of $7.12\pm 
0.11$~s \cite{GSI13}, we miss just a factor $(10.17\pm 0.26)/
(7.12\pm 0.11)=1.43\pm 0.06$ consistent with $\gamma=1.42$. 
However, this coincidence might well prove fortuitous. Minute departures 
from the assumed ${\vec\beta}\cdot\Delta{\vec p}_{Djj'}^{\,(P)}=0$ 
could lead to appreciable modifications of Eq.~(\ref{eq:DelE_L2}) 
for $\Delta E_{Djj'}^{(L)}$. 

To demonstrate the sensitivity of the result (\ref{eq:DelE_L2}) to decoherence 
effects arising from electron cooling, we suppress the kinetic energy 
contribution to $\Delta E_{Djj'}^{(L)}$ from the longitudinal components 
$p^{(L)}_{\parallel}$, keeping only the small transverse components 
$p^{(L)}_{\perp}=p^{(P)}\sin\theta_P$ which are less likely to be 
affected by the dominantly small-angle electron-cooling Coulomb scattering. 
Thus, retaining only $\Delta p_{\perp\,Djj'}^{(L)}$ contributions, we obtain  
\begin{equation} 
\Delta E_{Djj'}^{(L)}\approx \frac{p^{(P)}\sin\theta_P}{E_D^{(L)}}\,
\Delta p_{Djj'}^{(P)}\sin\theta_P=\frac{\Delta{m_{jj'}^2}}{2\gamma M_D}
{{\sin}^2\theta_P}\to \frac{\Delta{m_{jj'}^2}}{3\gamma M_D},  
\label{eq:DelE_L3}
\end{equation} 
where in the last step we averaged over ${\sin}^2\theta_P$. 
This $\Delta E_{Djj'}^{(L)}$ agrees within a factor of approximately 3 with 
that of (\ref{eq:DelE_L2}). The main message is that the relevant lab energy 
differences $\Delta E_{Djj'}^{(L)}$ remain of the same order as above, 
$\Delta{m_{\nu}^2}/2M$, and do not vanish.

\section{Conclusion} 
\label{sec:concl} 

In this work, we derived a nonvanishing oscillation frequency that a daughter 
ion moving in a storage ring should exhibit when disentangled by the action 
of the guiding external magnetic field from the electron-neutrino produced 
in the two-body EC decay of a parent ion coasting in the storage ring. This 
frequency invloves the neutrinos squared-mass difference $\Delta{m_{\nu}^2}$ 
divided by $2M_P$, commensurate with the reported GSI EC decay rate modulation 
frequency $\omega_{\rm EC}\sim 1~{\rm s}^{-1}$. It is specific to storage-ring 
motion, vanishing in rectilinear daughter-ion motion as discussed in 
Ref.~\cite{BK12}. The precise value of the period $T^{(L)}_{\rm osc}$ given 
under idealized conditions by Eq.~(\ref{eq:Tosc}) depends sensitively on 
the extent to which the motion of the daughter ion, incorporating its $D_j$ 
components, keeps coherent under given experimental constraints, primarily 
determined by the electron cooling system performance. In the unrealistic 
limit of full coherence, upon disregarding electron cooling, the related 
energy difference from Eq.~(\ref{eq:DelE_L1}) is $\Delta{m_{\nu}^2}/2E_{\nu}$ 
($\gg \Delta{m_{\nu}^2}/2M_P$) which leads to unobervable oscillations with 
a period of order 1~ms. 

Daughter-ion oscillations in storage-ring circular motion, as discussed 
in the present work, do not contradict any of the assertions made in past 
work~\cite{Giunti08,Kienert08,CGL09,Merle09,Flam10,Wu10,Pesh14}, namely that 
oscillations cannot show up in parent-ion $P$ decay rates or in daughter-ion 
$D$ appearance rates. In these works disentanglement is triggered by observing 
the EC decay, at which time the daughter ion $D$ and the undetected electron 
neutrino $\nu_e$ are entangled with each other, whereas the observation of $D$ 
in storage-ring circular motion guided by external EM fields occurs when $D$ 
is no longer entangled with $\nu_e$ from which it was disentangled at the 
$P\to D+\nu_e$ decay time. Therefore, none of these assertions hold for 
a storage-ring motion of the daughter ions. 

As long as GSI-like storage-ring EC decay experiments do not distinguish 
between observables pertaining to decay parent ions and observables associated 
with circulating daughter ions, daughter-ion oscillations will persist in 
giving rise to apparent modulation of EC decay rates, with amplitudes that 
depend on how the decay of parent ions $P$ and the subsequent storage-ring 
motion of daughter ions $D$ are both handled by the experimental setup 
and the final analysis of the experiment. The role of electron cooling 
in possibly preserving some of the coherence in the daughter-ion propagation 
at the GSI storage ring needs to be studied further, both experimentally and 
theoretically.

\section*{Acknowledgments} 
I would like to thank Fritz Bosch, Hans Feldmeier, Eli Friedman and Berthold 
Stech for making critical but useful remarks on several occasions during the 
preparation of previous versions. Stimulating discussions with Boris Kayser 
and Murray Peshkin during the recent EMMI Rapid Reaction Task Force workshop 
on Non-Exponential Two-Body Weak Decays at Jena, Germany, July 2014, 
are gratefully acknowledged. I also thank Fritz Bosch, Thomas Faestermann, 
Yuri Litvinov and Thomas St\"{o}hlker, who organized this workshop, 
for their kind hospitality and support.


\begin{thebibliography}{99}

\bibitem{GSI08} Litvinov, Yu.A.; Bosch, F.; Winckler, N.; {\it et al}. 
Observation of non-exponential orbital electron capture decays of 
hydrogen-like $^{140}$Pr and $^{142}$Pm ions. {\it Phys. Lett. B} {\bf 2008}, 
664, 162-168. 

\bibitem{GSI13} Kienle, P.; Bosch, F.; B\"{u}hler, P.; {\it et al}. 
(Two-Body-Weak-Decays Collaboration). High-resolution measurement of the 
time-modulated orbital electron capture and of the $\beta^+$ decay of 
hydrogen-like $^{142}$Pm ions. {\it Phys. Lett. B} {\bf 2013}, 726, 638-645. 

\bibitem{LBL08} Vetter, P.A.; Clark, R.M.; Dvorak, J.; Freedman, S.J.; 
Gregorich, K.E.; Jeppesen, H.B.; Mittelberger, D.; Wiedeking, M. Search 
for oscillation of the electron-capture decay probability of $^{142}$Pm. 
{\it Phys. Lett. B} {\bf 2008}, 670, 196-199. 

\bibitem{TUM09} Faestermann, T.; Bosch, F.; Hertenberger, R.; Maier, L; 
Kr\"{u}cken, R.; Rugel, G. Could the GSI decay rate oscillations be observed 
in a standard electron capture decay experiment? {\it Phys. Lett. B} 
{\bf 2009}, 672, 227-229. 

\bibitem{GSI14} Litvinov Yu.A. (spokesperson); Bosch, F. 
(for the Two-Body-Weak-Decays collaboration). ``Proposal for the continuation 
of the ESR-experiment E082" (July 2014, unpublished). See also the GSI 
technical report by Piotrovski, J.; Chen, X.; Litvinov, Yu.A.; Sanjari, M.S.; 
and the Two-Body-Weak-Decays collaboration. Performance of the ESR kicker 
magnet during E082. GSI Scientific Report 2014, {\bf 2015}, p.~429 
(doi:10.15120/GR-2015-1-FG-GENERAL-11). 

\bibitem{PDG} Abe, K.; {\it et al}. (Super-Kamiokande Collaboration). Solar 
neutrino results in Super-Kamiokande-III. {\it Phys. Rev. D} {\bf 2011}, 83, 
052010. 

\bibitem{IK09} Ivanov, A.N.; Kienle, P. Time Modulation of the $K$-Shell 
Electron Capture Decay Rates of H-like Heavy Ions at GSI Experiments. 
{\it Phys. Rev. Lett.} {\bf 2009}, 103, 062502. 

\bibitem{Gal10} Gal, A. Neutrino magnetic moment effects in electron-capture
measurements at GSI. {\it Nucl. Phys. A} {\bf 2010}, 842, 102-112.

\bibitem{Borexino08} Arpesella, C.; {\it et al}. (Borexino Collaboration). 
Direct Measurement of the $^7$Be Solar Neutrino Flux with 192 Days of 
Borexino Data. {\it Phys. Rev. Lett.} {\bf 2008}, 101, 091302. 

\bibitem{Giunti08} Giunti, C. Rates of processes with coherent production of 
different particles and the GSI time anomaly. {\it Phys. Lett. B} {\bf 2008}, 
665, 92-94. 

\bibitem{Kienert08} Kienert, H.; Kopp, J.; Lindner, M.; Merle, A. The GSI 
anomaly. {\it J. Phys. Conf. Ser.} {\bf 2008}, 136, 022049. 

\bibitem{CGL09} Cohen, A.G.; Glashow, S.L.; Ligeti, Z. Disentangling neutrino 
oscillations. {\it Phys. Lett. B} {\bf 2009}, 678, 191-196.  

\bibitem{Merle09} Merle, A. Why a splitting in the final state cannot explain 
the GSI-Oscillations. {\it Phys. Rev. C} {\bf 2009}, 80, 054616.

\bibitem{Flam10} Flambaum, V.V. Comment on ``Time Modulation of the $K$-Shell 
Electron Capture Decay Rates of H-like Heavy Ions at GSI Experiments". 
{\it Phys. Rev. Lett.} {\bf 2010}, 104, 159201. 

\bibitem{Wu10} Wu, J.; Hutasoit, J.A.; Boyanovsky, D.; Holman, R. Dynamics 
of disentanglement, density matrix, and coherence in neutrino oscillations. 
{\it Phys. Rev. D} {\bf 2010}, 82, 013006; 045027.  

\bibitem{Pesh14} Peshkin, M. Oscillating decay rate in electron capture and 
the neutrino mass difference. {\it Phys. Rev. C} {\bf 2015}, 91, 042501(R). 

\bibitem{NIMA04} Steck, M.; Beller, P.; Beckert, K.; Franzke, B.; Nolden, F. 
Electron cooling experiments at the ESR. {\it Nucl. Instrum. Methods A} 
{\bf 2004}, 532, 357-365. 

\bibitem{Dolgov97} Dolgov, A.D.; Morozov, A.Yu.; Okun, L.B.; Schepkin, M.G. 
Do muons oscillate? {\it Nucl. Phys. B} {\bf 1997}, 502, 3-18. For a related 
analysis of a different system, see Lowe, J.; Bassalleck, B.; Burkhardt, H.; 
Rusek, A.; Stephenson, G.J., Jr.; Goldman, T. No $\Lambda$ oscillations. 
{\it Phys. Lett. B} {\bf 1996}, 384, 288-292. 

\bibitem{BK12} Kayser, B. Neutrino Oscillation Physics. In {\it Proceedings 
of the International School of Physics ``Enrico Fermi"}; Bellini, G., 
Ludhova, L., Eds., IOS Press: Amsterdam, The Netherlands, 2012; pp.~1-14, 
and references listed therein [arXiv:1206.4325(hep-ph)]. 

\bibitem{Giunti14} For a recent discussion of decay-rate frame dependence, 
see Alavi, S.A.; Giunti, C. Which is the quantum decay law of relativistic 
particles? {\it Europhys. Lett.} {\bf 2015}, 109, 60001. 


\end{thebibliography}
\end{document}